\documentclass[sigconf,9pt]{acmart}
\AtBeginDocument{%
  \providecommand\BibTeX{{%
    \normalfont B\kern-0.5em{\scshape i\kern-0.25em b}\kern-0.8em\TeX}}}

\copyrightyear{2024}
\acmYear{2024}
\setcopyright{acmlicensed}\acmConference[DAC '24]{61st ACM/IEEE Design Automation Conference}{June 23--27, 2024}{San Francisco, CA, USA}
\acmBooktitle{61st ACM/IEEE Design Automation Conference (DAC '24), June 23--27, 2024, San Francisco, CA, USA}
\acmDOI{10.1145/3649329.3657353}
\acmISBN{979-8-4007-0601-1/24/06}

\usepackage{listings}
\usepackage[frozencache,cachedir=.]{minted}
\usepackage{tcolorbox}
\usepackage{caption}
\usepackage{subcaption}
\usepackage{tabularx}
\begin{document}

\title{\NAME: Automatically Fixing RTL Syntax Errors with Large Language Models}


\author{YunDa Tsai}
\authornote{equal contribution}
\email{yundat@nvidia.com}
\affiliation{%
  \institution{NVIDIA}
  \country{}
}

\author{Mingjie Liu}
\authornotemark[1]
\email{mingjiel@nvidia.com}
\affiliation{%
  \institution{NVIDIA}
  \country{}
}

\author{Haoxing Ren}
\email{haoxingr@nvidia.com}
\affiliation{%
  \institution{NVIDIA}
  \country{}
}

% \affiliation{%

% \affiliation{%

% \affiliation{%

% \affiliation{%

% \affiliation{%

\renewcommand{\shortauthors}{Tsai, et al.}
\def\bibfont{\footnotesize}
\newcommand{\NAME}{\textbf{RTLFixer}}

\begin{abstract}
  This paper presents \NAME, a novel framework enabling automatic syntax errors fixing for Verilog code with Large Language Models (LLMs). Despite LLM's promising capabilities, our analysis indicates that approximately 55\% of errors in LLM-generated Verilog are syntax-related, leading to compilation failures. To tackle this issue, we introduce a novel debugging framework that employs Retrieval-Augmented Generation (RAG) and ReAct prompting, enabling LLMs to act as autonomous agents in interactively debugging the code with feedback. This framework demonstrates exceptional proficiency in resolving syntax errors, successfully correcting about 98.5\% of compilation errors in our debugging dataset, comprising 212 erroneous implementations derived from the VerilogEval benchmark. Our method leads to 32.3\% and 10.1\% increase in pass@1 success rates in the VerilogEval-Machine and VerilogEval-Human benchmarks, respectively.
  The source code and benchmark are available at \url{https://github.com/NVlabs/RTLFixer}.
\end{abstract}

\maketitle
\pagestyle{plain}

\section{Introduction}
% intro
%Verilog code generation using Large Language Models (LLMs) represents a significant advancement in hardware design automation. Previous works like VeriGen~\cite{thakur2023verigen} and VerilogEval~\cite{liu2023verilogeval} have concentrated on zero-shot code generation. However, like many complex programming tasks, producing error-free code in a single attempt is challenging, with a high likelihood of errors. Given that even human coders often require multiple attempts to get it right, there is a clear need for effective debugging and refinement capabilities in LLM-generated Verilog code.

Large language models (LLMs) present great promise in automating hardware design, especially in their capacity to understand design intentions and produce Verilog code from natural language~\cite{liu2023chipnemo}. Recent efforts, such as VeriGen~\cite{thakur2023verigen} and VerilogEval~\cite{liu2023verilogeval}, have primarily focused on zero-shot code generation. However, akin to numerous complex programming tasks, generating flawless code in a single attempt poses a significant challenge, with a high likelihood of errors. Consequently, there exists a clear need for robust debugging and refinement capabilities in Verilog code generated by Large Language Models. This need arises from that like human programmers, achieving precision often requires multiple iterations.

% motivation
%The debugging process can typically be broken down into three phases: 1) Fixing syntax errors, 2) Obtaining simulation results, and 3) Refining the code based on feedback. Our analysis reveals that 60\% of LLM-generated codes contain errors, with syntax errors constituting 56\% of these. These syntax errors are critical roadblocks, preventing successful compilation and simulation, thus hindering the debugging cycle. Addressing syntax errors is therefore crucial, as it significantly reduces the effort required to select a valid code sample from numerous alternatives.
Most importantly, it is evident that Large Language Models encounter challenges in generating fully syntactically correct Verilog code.  Surprisingly, our analysis reveals that a substantial 55\% of the errors generated by LLMs for Verilog code are comprised of syntax errors, surpassing the occurrence of logic errors detected through simulation.
%{\color{blue} Show the "most common" example of syntax error of LLM generated code. State that fixing it could lead to correct implementation.}
Rectifying syntax errors not only enhances the overall accuracy of LLM-generated code but also holds the potential to alleviate manual efforts for human engineers engaged in Verilog coding. The recognition and mitigation of syntax errors stand as imperative steps, not only for refining LLM capabilities but also for streamlining the coding process for human practitioners in the domain of Verilog development.
%{\color{blue} can we show examples of syntax errors in introduction as motivating examples}

% research problem
%{\color{blue}  talk about ReAct , and those related works are using ReAct techniques, right? }
%Related works, such as SELF-DEBUG~\cite{chen2023teaching}, SELFEVOLVE~\cite{jiang2023selfevolve} and AutoChip~\cite{thakur2023autochip}, approach automatic debugging by returning feedback to the LLM and relying on its inherent knowledge for code refinement. However, these methods primarily increase the likelihood of generating syntactically correct code, but do not necessarily ensure the complete elimination of syntax errors. This raises the question: how can we fully rectify syntax errors in LLM-generated Verilog code?

\begin{figure}[t]
\centering
  \centering
  \includegraphics[width=\linewidth]{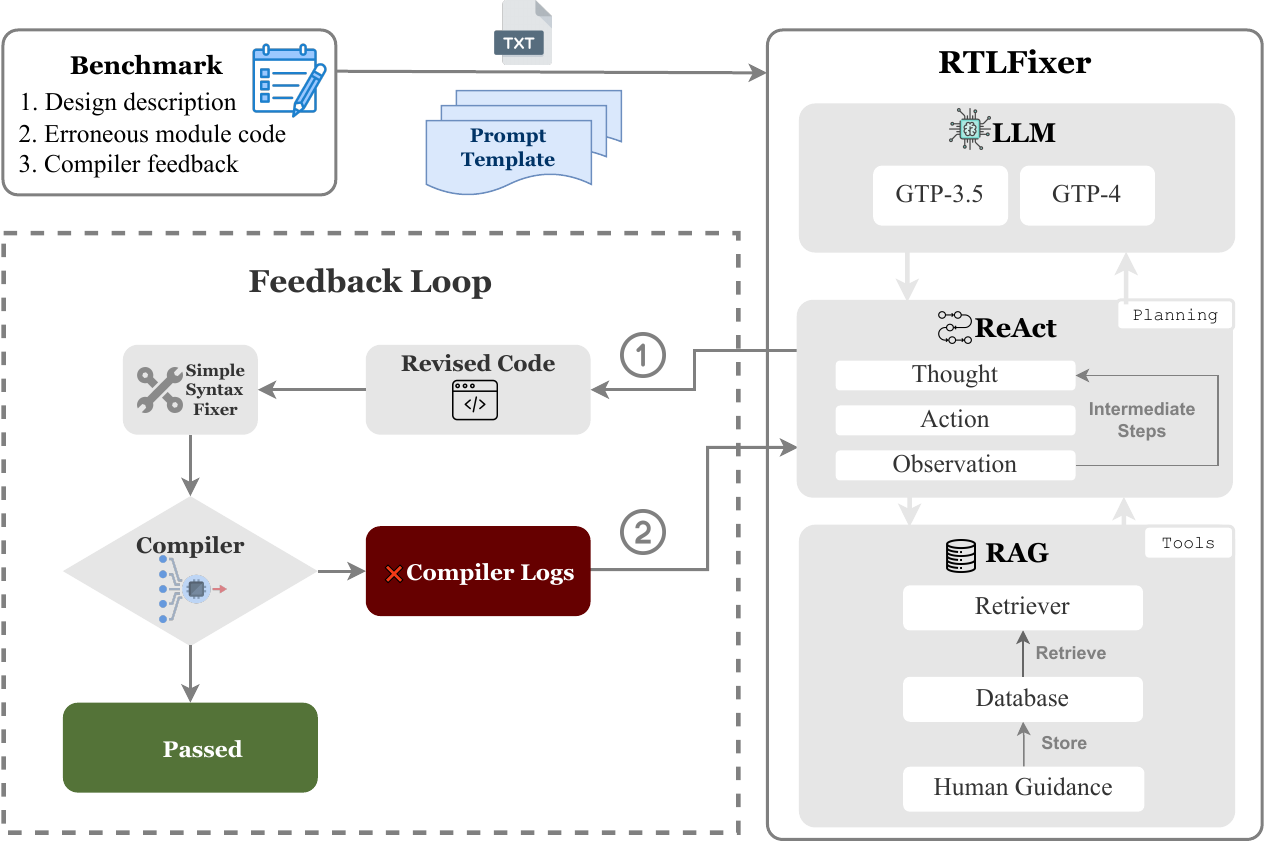}
  \caption{Overview of \NAME. The Autonomous Language Agent fixes the syntax error via a feedback loop. ReAct handles the iterative code refinement with intermediate reasoning and action steps. Human expert guidance is incorporated through RAG. 
  }
  \label{fig:debugging-framework}
\end{figure}

Despite such challenges, LLMs have showcased remarkable capabilities in reasoning and enhancing action plans to address exceptions. The ReAct framework~\cite{yao2022react} integrates reasoning and action synthesis in language models, demonstrating LLM's capability to engage in reasoning processes and refine decision-making through interactive feedback. Similarly, SelfDebug~\cite{chen2023teaching} and SelfEvolve~\cite{jiang2023selfevolve} illustrate the model's ability to self-identify mistakes by scrutinizing execution results and articulating generated code in natural language. 
%SelfEvolve~\cite{jiang2023selfevolve} further advances this line of research by tasking LLMs with debugging generated code, incorporating feedback from the code execution interpreter. 
It is essential to highlight that prior works do not explicitly address the correction of syntax errors and primarily center on improving the accuracy of generated code, particularly in Python, where language models excel syntactically.

%{\color{blue} briefly talk about RAG benefits}

On the other hand, Large Language Models are acknowledged for their inclination to produce factual errors, a phenomenon termed hallucination~\cite{Ji_2023}.
%, attributed to the unvetted and non-fact-checked nature of the extensive training corpus derived from the internet. 
%Specifically in coding contexts, LLMs have exhibited the generation of inaccurate API calls for Python data science packages~\cite{lai2022ds1000}. 
To mitigate this challenge, the Retrieval-Augmented Generation (RAG) paradigm~\cite{lewis2021retrievalaugmented} has been introduced, integrating retrieval mechanisms to improve the precision of generated content by incorporating information from external knowledge sources.%, greating reducing hallucinations. 

In this paper, we introduce \NAME, a innovative debugging framework that utilizes LLMs as autonomous language agents in conjunction with RAG. We exclusively focus on addressing the challenge of rectifying syntax errors in RTL Verilog code—an essential problem with potential benefits for both LLMs and human engineers. 
As shown in Figure~\ref{fig:debugging-framework}, our framework combines established human expertise stored in a retrieval database for correcting syntax errors while simultaneously harnessing the capabilities of LLMs as autonomous agents for reasoning and action planning (ReAct). Through incorporating human expertise, our approach provides explicit guidance and explanations when LLMs face challenges in error correction. The stored compiler messages and human expert guidance function as a persistent external non-parametric memory database, enhancing results through RAG. By empowering LLMs with ReAct, LLMs serve as autonomous agents adept at strategically planning intermediate steps for iterative debugging. %We establish a feedback loop incorporating message logs from Verilog code compilers, utilizing Large Language Models (LLMs) as autonomous agents that iteratively debug solutions through active interactions with the compiler.  
We also create VerilogEval-syntax, a Verilog syntax debugging dataset, derived from VerilogEval~\cite{liu2023verilogeval}, containing 174 erroneous implementations.

Our contributions are summarized as follows:
\begin{itemize}
    \item Our framework demonstrates an impressive 98.5\% success rate in resolving syntax errors, resulting in a noteworthy 32.3\% and 10.1\% improvement in the pass@1 metrics achieved solely by addressing syntax errors in VerilogEval-Machine and VerilogEval-Human benchmarks, respectively. 
    \item Our framework also improves the syntax success rate from 73\% to 93\% on the RTLLM benchmark~\cite{lu2023rtllm}, demonstrating the generalizability of this approach. 
    \item Compared to One-shot generation, ReAct enhances syntax success rates by 25.7\%, 26.4\%, and 31.2\% with iterative feedback from Simple, iverilog, and Quartus, respectively.
    \item RAG with human guidance significantly improves syntax success rates, up to 31.2\% and 18.6\% with feedback from Quartus for One-shot and ReAct prompting, respectively.
    
\end{itemize}

The remainder of this paper is structured as follows. In Section~\ref{sec:prelim}, we present preliminary works on LLMs for Verilog code generation, ReAct, and RAG. Our debugging framework \NAME~is elucidated in Section~\ref{sec:method}, where we empower LLMs as autonomous agents with ReAct and innovatively provide human guidance through RAG. Section~\ref{sec:experiments} details our experimental results, showcasing the effectiveness of our method in correcting syntax errors and improving the pass rate. Finally, Section~\ref{conclusion} summarizes and concludes the paper.

%As illustrated in Figure~\ref{fig:debugging-framework}, the framework incorporates human expertise to offer explicit guidance and explanations during instances where the LLM encounters challenges in error correction. These code segments, compiler logs, and human guidance are stored for future retrieval, essentially serving as a persistent external non-parametric memory for the LLM. Furthermore, we establish a feedback loop that captures logs from Verilog code compilers, where LLMs empowered by reasoning and enhanced action planning (ReAct)~\cite{yao2022react} serve as autonomous agents, strategically planning intermediate steps for iterative debugging. The agent autonomously prompts itself to revise code, generate error explanations, and retrieve past solutions until the error is rectified or a maximum iteration threshold is reached. Our framework demonstrates an impressive 98.5\% success rate in resolving syntax errors, resulting in a noteworthy 5\% improvement in the overall pass@5 metric achieved solely by addressing syntax errors.

%{\color{blue} need to write a contribution list}

\section{Preliminaries}
\label{sec:prelim}
In this section, we begin by briefly exploring the advancements and applications of LLMs for Verilog code generation in Section~\ref{sec:pre-llm-code-gen}. We then delve into the synthesis of reasoning and action in LLMs, elaborated in Section~\ref{sec:pre-reason-action}. Finally, we discuss Retrieval-Augmented Generation in Section~\ref{sec:pre-rag}.

\subsection{LLMs for Verilog Code Generation}
\label{sec:pre-llm-code-gen}
% Talk about all related works on LLMs for code generation here. can also include broad code generation.

Large Language Models exhibit the capability to generate code, with Codex~\cite{chen2021evaluating} standing as an early exemplar. GitHub Copilot~\cite{friedman2021introducing}, building upon such groundwork, played a crucial role in pioneering LLM-based code completion engines, contributing significantly to the domains of auto-completion and conversational code generation. DAVE~\cite{pearce2020dave} emerged as an early study of LLM tailored for hardware design. VeriGen~\cite{thakur2023verigen} further expanded the dataset scope and experimented with open-sourced models. Following suit, Chip-Chat~\cite{blocklove2023chip}, leveraging GPT-4, demonstrated the extensive potential of LLMs in collaboratively generating processors and other hardware designs. In parallel, benchmarks such as VerilogEval~\cite{liu2023verilogeval} and RTLLM~\cite{lu2023rtllm} played a pivotal role in advancing the application of LLMs in Verilog code generation.

\subsection{Reasoning and Action Synthesis of LLMs}
\label{sec:pre-reason-action}
% Talk about LLMs for reasoning. Include works such as chain of thoughts etc.
% Talk about LLMs for autonomous agents. Action.
% Talk about ReAct in DETAILS. Integrate the 2? Why is ReAct better.
%Large Language Models (LLMs) have excelled in both reasoning and acting in various tasks. For reasoning, techniques like chain-of-thought prompting enable LLMs to decompose complex problems into a series of logical steps, significantly enhancing their problem-solving abilities. In terms of acting, LLMs, exemplified by models such as WebGPT~\cite{nakano2021webgpt}, Toolformer~\cite{schick2023toolformer} and ToolLLM~\cite{qin2023toolllm}, are adept at interactive decision-making and creating action plans, often utilizing digital tools effectively.

%The ReAct~\cite{yao2022react} framework represents a significant leap in LLM capabilities by integrating these two aspects. It allows LLMs to generate both reasoning traces and specific actions, facilitating dynamic interaction with external information sources. This integration not only improves LLMs' performance in complex tasks but also makes them more reliable and versatile as autonomous agents, capable of more accurate and context-aware responses.

Large Language Models (LLMs) have demonstrated proficiency in both reasoning and planning across various tasks. In terms of reasoning, methods like chain-of-thought~\cite{wei2023chainofthought} empower LLMs to break down intricate problems into logical steps, significantly enhancing their problem-solving abilities. LLMs also excel in interactive decision-making and formulating action plans, effectively leveraging digital tools, as shown in works such as ToolLLM~\cite{qin2023toolllm}.
%In the realm of action planning, LLMs, as seen in models like WebGPT~\cite{nakano2021webgpt}, Toolformer~\cite{schick2023toolformer}, and ToolLLM~\cite{qin2023toolllm}, excel in interactive decision-making and formulating action plans, effectively leveraging digital tools.

The ReAct~\cite{yao2022react} framework represents a notable advancement in LLM capabilities by seamlessly integrating reasoning and action planning. This framework enables LLMs to generate both reasoning traces and specific actions, facilitating dynamic interaction with external information sources. This integration not only enhances LLM's performance in complex tasks but also renders them more reliable and versatile as autonomous agents, capable of delivering more accurate and context-aware responses.

\subsection{Retrieval-Augmented Generation}
\label{sec:pre-rag}
% Talk about retrieval-augmented generation. This also should be slightly in detail.

Retrieval-Augmented Generation (RAG)~\cite{lewis2021retrievalaugmented} represents a significant advancement in addressing the limitations of Large Language Models when handling knowledge-intensive tasks. LLMs despite containing extensive factual knowledge, often encounter challenges in accessing and effectively manipulating this information. RAG leverages a combination of LLMs, which serve as parametric memory, and an external knowledge base, such as Wikipedia, functioning as non-parametric memory. This unique approach allows RAG to access and retrieve relevant documents or passages from the external knowledge base based on the input query. Consequently, this enriches the context available to the text generator, resulting in outputs that exhibit improved accuracy and factual consistency. In the context of code generation, several works such as ReACC~\cite{lu2022reacc} and RepoCoder~\cite{zhang2023repocoder} have successfully harnessed the capabilities of RAG to enhance the code generation proficiency of LLMs, showcasing its transformative potential.

\section{\NAME: Resolving Syntax Error with LLM Agents and Retrieval}
\label{sec:method}

In this section, we explain the details of \NAME, which utilizes Autonomous Language Agents enhanced with ReAct and Retrieval-Augmented Generation (RAG). The framework's structure is outlined in Figure~\ref{fig:debugging-framework} (Section~\ref{sec:method:overview}), and the application of ReAct is thoroughly discussed in Section~\ref{sec:method:react}. Furthermore, Section~\ref{sec:method:rag} explores the integration of human expert guidance using RAG. Finally, we explain the curation process for our VerilogEval-syntax error dataset.

\subsection{Overview of \NAME}
\label{sec:method:overview}
\NAME~comprises an LLM for code generation, RAG for accessing human expert guidance, and ReAct for improved task decomposition, tool use, and planning. Our approach starts by formulating an input prompt integrating a benchmark dataset problem into a template, followed by the agent utilizing RAG and ReAct, revising erroneous Verilog code. If syntax errors persist, error logs from the compiler as well as retrieved human guidance from the database are provided as feedback. This interactive debugging loop can be repeated multiple times until all errors are resolved. %The input prompt includes system prompts, natural language problem descriptions, incorrect Verilog implementations, and compiler logs. %An example is shown in Figure~\ref{fig:input-prompt}. 

% \begin{figure}[h!]
% \centering
%   \centering
%   \includegraphics[width=\linewidth]{Figs/input-prompt.drawio.pdf}
%   \caption{An exemplar of input prompt template.}
%   \label{fig:input-prompt}
% \end{figure}

\subsection{Reasoning and Action Planning through ReAct Iterative Prompting}
\label{sec:method:react}

We enable Large Language Models to function as autonomous agents for reasoning and action planning through the \textbf{ReAct} prompting mechanism~\cite{yao2022react}. In ReAct, LLMs generate both reasoning traces and task-specific actions in an interleaved manner. The input prompt, along with the ReAct instruction prompt, is provided to an LLM. Subsequently, the LLM initiates the generation of ReAct steps, each consisting of Thought, Action, and Observation components. An example of a ReAct instruction prompt is depicted in Figure~\ref{fig:react-instruct}, while Figure~\ref{fig:react-steps} illustrates the self-prompting process, showcasing the intermediate steps within each iteration of ReAct.

During this process, the LLM prompts itself for thoughts on how to address the error and selects the next action. Potential actions include generating an explanation for the error, searching for a solution in the human expert guidance database, revising the code, and submitting the revised code to the compiler, among other possibilities. The output of the chosen action becomes the observation in the prompt. The agent continues prompting until the compilation is successful, selecting the Finish action to output the final response. If unsuccessful, the process iterates up to n times, where n is a user-selected hyperparameter. Our objective is to assess the effectiveness of a fully automated feedback-driven solution.

We employ \textbf{One-shot} prompting, illustrated in Figure~\ref{fig:input-prompt}, as the baseline for comparing with ReAct. One-shot prompting involves only a single-turn of feedback from the compiler message, but it excludes ReAct prompting, which decomposes syntax fixing through reasoning and action planning, and the iterative multi-round interactions with the compiler.

\begin{figure}[h]

    % \includegraphics[width=\linewidth]{Figs/react-prompt.drawio.pdf}
    % \vspace{-0.4cm}
    
    \begin{subfigure}[b]{0.47\textwidth}

        \begin{tcolorbox}[width=1.0\linewidth, halign=left, colframe=black, colback=white, boxsep=0.01mm, arc=1.5mm, left=2mm, right=2mm, boxrule=0.5pt]\footnotesize
    
        \textbf{System Prompt:}\\ 
        Implement the Verilog module based on the following description. Assume that signals are positive clock/clk edge triggered unless otherwise stated.
        
        \vspace{0.1in}
        \textbf{Problem Description:}\\  
        Implement the Verilog module based on the following description. Assume that signals are positive clock/clk edge triggered unless otherwise stated.\\
        Given an 8-bit input vector [7:0], reverse its bit ordering.\\
        \begin{minted}[breaklines]{verilog}
module top_module (
input [7:0] in,
output [7:0] out
);
        \end{minted}
    
        \vspace{0.1in}
        \textbf{Erroneous Implementation:}\\  
        \begin{minted}[breaklines]{verilog}
assign {out[0],out[1],out[2],out[3],out[4],out[5],out[6],out[8]} = in;
endmodule
        \end{minted}

        \vspace{0.1in}
        \textbf{Feedback:}\\  
        main.v:5: error: Index out[8] is out of range.\\
        1 error(s) during elaboration.\\
        
        \end{tcolorbox}
        \vspace{-0.1in}
        \caption{One-shot Prompt Template}
        \label{fig:input-prompt}
    \end{subfigure}
    
    %\vspace{0.1in}
    
    \begin{subfigure}[b]{0.47\textwidth}

        \begin{tcolorbox}[width=1.0\linewidth, halign=left, colframe=black, colback=white, boxsep=0.01mm, arc=1.5mm, left=2mm, right=2mm, boxrule=0.5pt]\footnotesize
    
        Solve a task with interleaving Thought, Action, Observation steps.\\
        Thought can reason about the current situation, and Action can be the following types:\\
        (1) Compiler[code], which compiles the input code and provide error message if there is syntax error.\\
        (2) Finish[answer], which returns the answer and finished the task.\\
        (3) RAG[logs], input the compiler log and retrieve expert solutions to fix the syntax error.\\
        
        \end{tcolorbox}
        \vspace{-0.1in}
        \caption{ReAct Instruction (System Prompt)}
        \label{fig:react-instruct}
    \end{subfigure}

    %\vspace{0.1in}
    \begin{subfigure}[b]{0.47\textwidth}

        \begin{tcolorbox}[width=1.0\linewidth, halign=left, colframe=black, colback=white, boxsep=0.01mm, arc=1.5mm, left=2mm, right=2mm, boxrule=0.5pt]\footnotesize
    
        \textbf{Question:}\\ 
        What is the syntax error in the given Verilog module implementation and how to fix it?
        
        \vspace{0.1in}
        \textbf{Thought 1:}\\  
        Implement a finite state machine that searches for message boundaries in a byte stream.\\
        \textbf{Action 1: Compiler}\\  
        \begin{minted}[breaklines]{verilog}
    always @ (state or in) begin
    case (state)
     4'b0001: begin
        \end{minted}
    
        \textbf{Observation 1:}\\  
        main.v:15: error: out is not a valid l-value in top\_module.\\
    
        \textbf{Thought 2:}\\  
        The code has a syntax error because the "out" signal was declared as a wire instead of a register. I need to correct this and run the compilation.\\
        \textbf{Action 2: RAG[..not a valid l-value..]}\\
        Use assign statements instead of always block if possible.\\
        \centering ... \\
        \raggedright
        \textbf{Action n: Finish}\\
        The revised code is as follow: ...\\
        
        \end{tcolorbox}
        \vspace{-0.1in}
        \caption{ReAct Steps with Thought-Action-Observation.}
        \label{fig:react-steps}
    \end{subfigure}

  \caption{Prompts used for ReAct. (a) shows the One-shot prompting template with feedback message. (b)-(c) demonstrate the example where LLMs serve as autonomous agents with ReAct to decompose syntax fixing problems with reasoning and planning.}
  \label{fig:prompt}
\end{figure}

\begin{figure}[ht]
\begin{tcolorbox}[width=1.0\linewidth, halign=left, colframe=black, colback=white, boxsep=0.01mm, arc=1.5mm, left=2mm, right=2mm, boxrule=0.5pt]\footnotesize

\textbf{Compiler Logs:}\\ 
Object `clk' is not declared. Verify the object name is correct. If the name is correct, declare the object.

\textbf{Human Expert Guidance:}\\ 
Check if `clk' is an input. If not, and if `clk' is used within the module, make sure the name is correct. If it's meant to trigger an `always' block, replace `posedge clk' with `*'.

\vspace{0.2in}
\textbf{Compiler Logs:}\\ 
Index cannot fall outside the declared range for vector\\

\textbf{Human Expert Guidance:}\\ 
Carefully examine the index values to prevent encountering `index out of bound' errors in your code. When utilizing parameters for indexing, try to use binary strings for performing the indexing operation instead.
\end{tcolorbox}
\caption{Examples of common error categories that LLM constantly could not solve and the corresponding human expert guidance in the retrieval database.}
\label{fig:rag-feedback}
\end{figure}

\subsection{Retrieval Augmented Generation (RAG)}
\label{sec:method:rag}

We leverage Retrieval-Augmented Generation (RAG), a potent technique that notably enhances Large Language Models' capabilities by incorporating human expert guidance through a retriever. A key distinction from traditional RAG lies in our curated database, enriched with human instructions and demonstrations.

The retrieval database curation process involves a meticulous procedure of categorizing syntax errors and developing instructions and demonstrations for syntax error resolution. In the initial step, we categorize various syntax errors into groups using error number tags provided by compilers (such as Quartus) in the compiler logs. During the manual inspection of LLM's struggle cases, it becomes evident that ambiguous error messages present a significant challenge, impeding the model's error resolution capabilities. The inclusion of clear instructions, and demonstrations of possible solutions enables the LLM to adeptly address errors. To facilitate this, human experts offer detailed explanations for compiler logs, serving as human expert guidance. An illustrative example of common errors is showcased in Figure~\ref{fig:rag-feedback}. Subsequently, all compiler logs, error code segments, and corresponding human guidance undergo systematic storage in the database for future retrieval.

We integrate the human guidance and demonstration database with Large Language Models (LLMs) using Retrieval-Augmented Generation (RAG). RAG enables the retrieval of pertinent documents or data from a source, utilizing this information as context for the original input prompt. This approach allows the language model to access the latest information without the need for retraining, proving especially valuable in enhancing the model's capacity to generate more accurate and reliable outputs. Leveraging RAG further ensures that the language model has access to the most current and relevant information, including compiler logs and human guidance, facilitating effective error resolution.

Figure~\ref{fig:rag-feedback} illustrates two common error categories along with a demonstration of compiler logs and corresponding human  guidance. For this task, common retrievers such as pattern-matching, fuzzy search, or similarity search with a vector database are suitable. In our experiments, we opted for an exact match to error tags for simplicity, given the limited number of error cases.
We collected 7 common error categories with 30 entries for iverilog and 11 common error categories with 45 entries for Qaurtus in total.
% {\color{blue} How many entries in the dataset in total, how many for iverilog v.s. Quartus.}

% \begin{figure}[h!]
% \centering
%   \centering
%   \includegraphics[width=\linewidth]{Figs/rag-feedback.pdf}
%   \caption{7 common error categories that LLM constantly could not solve and the corresponding human guidance.}
%   \label{fig:rag-feedback}
% \end{figure}

\subsection{Debugging Dataset}
We created a novel benchmark dataset, VerilogEval-syntax, based on the VerilogEval benchmark~\cite{liu2023verilogeval}. This dataset comprises flawed code implementations sourced from the VerilogEval problem set. Each entry includes the original problem description and erroneous implementation containing syntax errors.
%Additionally, Fig.\ref{fig:error-stat} presents the statistics of errors in code generation using the VerilogEval benchmark, highlighting that syntax errors constitute a significant 56\% of the total errors produced by LLMs in Verilog code generation. 

The dataset curation includes sampling, filtering, and clustering. Code samples were selected from VerilogEval problems using One-shot and ReAct prompting methods with \textit{gpt-3.5-turbo} model, retaining only error-inducing samples. In the filtering phase, we focus on code with compile errors and use the following processing and filtering criteria: extraction of code from markdown blocks, validation of module statements, and removal of samples with extraneous language or empty module bodies. The final step involved clustering using DBSCAN~\cite{schubert2017dbscan} with Jaccard distance~\cite{niwattanakul2013using}, grouping similar implementations to select representative examples while ensuring a diverse representation of syntax errors. This results in a total of 212 erroneous implementations in the dataset.

\section{Experiments}
\label{sec:experiments}
In this section, we first present the evaluation metrics in Section~\ref{sec:exp:eval}, followed by our primary findings showcased in Section~\ref{sec:exp:main}. Within this section, we show the performance improvements and the impact of ReAct and RAG. Finally, Section~\ref{sec:exp:ablation} details ablation studies on the quality feedback message and LLM.

\noindent\textbf{Setup:} We conduct all experiments with GPT-3.5 as the LLM through OpenAI APIs~\cite{openai}, except for the ablation experiment on different LLM. We specifically used \textit{gpt-3.5-turbo-16k-0613}. 
%Note that we also utilize the OpenAI function calling feature to restrict the LLM response into certain format.
% The default compiler will be Quartus except for the ablation experiment on different compilers.
A simple rule-based syntax fixer is applied to every LLM-generated verilog code, which avoids simple errors such as misplaced timescale derivatives. 
%This is for avoiding simple errors and facilitate the experiment without wasting budgets for LLM sampling.
In all experiments, we set the sampling temperature to 0.4. %parameter consistently at 0.4.
%Note that LLMs with identical prompts and the same temperature settings can still yield different outputs.
For ReAct prompting, we restrict the LLM to a maximum of 10 iterations of Thought-Action-Observation, where Action might involve interactions with the compiler. We consider the syntax error resolved if any of the generated code passes. 
To limit test variance, we repeat each experiment 10 times and report the average. 
%For our ReAct prompting, the number of feedback iterations is capped at 10. 

% \textbf{LLM}: All the experiments are conducted with gpt-3.5-turbo as the LLM through OpenAI APIs~\cite{openai}, except for the ablation experiment on different LLM. We specifically used the `gpt-3.5-turbo-16k-0613` version. Note that we also utilize the OpenAI function calling feature to restrict the LLM response into certain format.

% \noindent\textbf{Compiler}: The default compiler will be Quartus except for the ablation experiment on different compilers.

% \noindent\textbf{Simple Syntax Fixer}: A simple rule-based syntax fixer is applied to every LLM generated verilog code. This is for avoiding simple errors and facilitate the experiment without wasting budgets for LLM sampling. The rules are as follow:
% \begin{enumerate}
%     \item Validate if the module header defined in the problem spec exists.
%     \item Validate if `endmodule` statement is at the end to form a complete module block.
%     \item Validate if `timescale` derivative is misplaced.
% \end{enumerate}

% \noindent\textbf{Hyper Parameters}:
% In our experiments, we set the temperature parameter consistently at 0.4.
% Note that LLMs with identical prompts and the same temperature settings can still yield different outputs. Beyond this, all the other parameters
% To ensure robustness, each experiment is repeated 10 times, and the outcomes are averaged. For our ReAct prompting, the number of feedback iterations is capped at 10. Additionally, we use the default parameters for the LLMs.

\subsection{Evaluation Metric}
\label{sec:exp:eval}
\noindent\textbf{Compile Fix rate}: To demonstrate the debugging capability of our method, we calculate the expectation fix rate, with $c$ as the number of fixed samples out of all $n=10$ samples. 

%For Direct prompting, we generate $n=10$ samples for each problem and report the average. For one single chain of ReAct, we restrict the LLM to a maximum of 10 feedback iterations involving interactions with the compiler. We consider the syntax error resolved if any of the generated code passes. We create 10 ReAct chains for each problem and report the average, where each chain comprises up to 10 iterations. 

\begin{align}
    \textrm{fix rate} = \underset{problems}{\mathbb{E}} \left[ \frac{c}{n} \right]
\end{align}
% \item Compile per fix: We calculate the number of code revision needed to fix the compile error for iterative methods such as ReAct to evaluate the efficiency of iterations.
\noindent\textbf{Functional Correctness}: We follow recent work in directly measuring code functional correctness with simulation through pass@k metric~\cite{chen2021evaluating}, where a problem is considered solved if any of the k samples passes the tests. We use the unbiased estimator as follow and ensure $n=20$ is sufficiently large:
\begin{align}
    \textrm{pass@k} = \underset{problems}{\mathbb{E}} \left[ 1-\frac{{n-c \choose k}}{{n \choose k}} \right]
\end{align}

\subsection{Main Results}

\begin{table}[ht]
    \centering
    \footnotesize
    \begin{tabular}{ |lc|c|cc|c| }
     %\hline
     %\multicolumn{6}{|c|}{VerilogEval-syntax} \\
     \hline
     Prompt & RAG & Simple & iverilog & Quartus & GPT-4  \\
     \hline
     One-shot & w/o & 0.414 & 0.536 & 0.587 & 0.91  \\ 
     ~ & w/ & - & 0.800 & 0.899 & 0.98  \\
     \hline
     ReAct & w/o & 0.671 & 0.731 & 0.799 & 0.92 \\ 
     ~ & w/ & - & 0.820 & 0.985 & 0.99   \\ 
     \hline
    \end{tabular}
    \caption{Fix rate for One-shot vs. ReAct, w/ and w/o RAG, ablation on feedback quality and LLMs on VerilogEval-syntax.}
    \label{tab:compile}
\end{table}

\begin{table}[ht]
    \footnotesize
    \begin{tabular}{ |ll|cc|cc| }
    \hline
    Dataset & Set & \multicolumn{2}{c|}{pass@1} & \multicolumn{2}{c|}{pass@5} \\
    VerilogEval & ~ & original & fixed & original & fixed \\
    \hline
    Human & All & 0.267 & 0.368 & 0.458 & 0.506 \\
    ~ & easy & 0.521 & 0.666 & 0.808 & 0.847 \\
    ~ & hard & 0.053 & 0.120 & 0.164 & 0.221 \\
    \hline
    Machine & All & 0.467 & 0.799 & 0.691 & 0.891 \\
    ~ & easy & 0.568 & 0.833 & 0.782 & 0.892 \\
    ~ & hard & 0.367 & 0.771 & 0.601 & 0.890 \\
    \hline
    \end{tabular}
    \caption{Pass@k for simulation pass rate on VerilogEval dataset after fixing syntax errors.}
    \label{tab:simulation}
\end{table}

\begin{figure}[ht]
\centering
  \centering

 %  \begin{subfigure}[b]{0.25\textwidth}
 %     \centering
 %     \includegraphics[width=\textwidth]{Figs/VerilogEval-Human-pie-before.pdf}
 % \end{subfigure}
 % \hfill
 % \begin{subfigure}[b]{0.22\textwidth}
 %     \centering
 %     \includegraphics[width=\textwidth]{Figs/VerilogEval-Human-pie-after.pdf}
 % \end{subfigure}

  % \vspace{-0.5cm}
  \includegraphics[width=\linewidth]{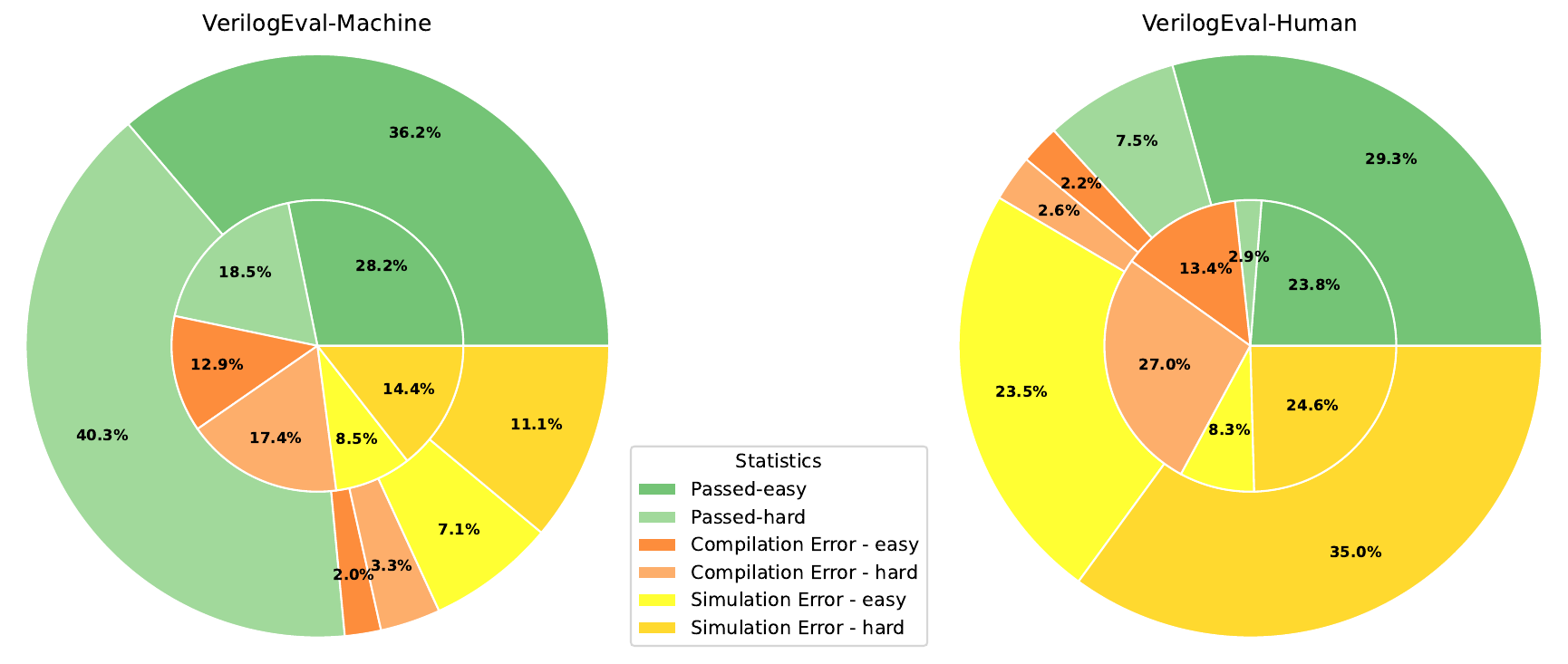}
  \caption{VerilogEval pass@1 results prior (inner) and post (outer) syntax error fixing with \NAME.}
  \label{fig:error-stat}
  \vspace{-0.1in}
\end{figure}

\label{sec:exp:main}
The main results in Table~\ref{tab:compile} show the effectiveness of ReAct and RAG which each provided performance gain in a large margin. We note that One-shot generation includes only a single-turn interaction with either simple or compiler message as feedback. Table~\ref{tab:simulation} showed the improvement of pass@\{1,5\} on VerilogEval dataset after fixing syntax errors with visualizations in Figure~\ref{fig:error-stat}. The VerilogEval-Human benchmark statistics show that syntax errors constitute a significant 55\% of errors in GPT-3.5 generated Verilog code, surpassing simulation errors. With just addressing syntax errors using our approach (shown in the inner circle), the pass rate increases from 26.7\% to 36.8\%. Table~\ref{tab:rtllm} shows that our approach can generalize across different benchmarks. We further detail our findings below.

\noindent\textbf{Impact of ReAct:} ReAct significantly outperforms One-shot generation. ReAct's ability to iteratively revise code with reasoning and planning results in superior performance. Moreover, even without explicit feedback from the compiler (Simple), the intermediate reasoning steps similar to chain-of-thought can still bring considerable improvements from 41.4\% to 67.1\%. When compared with One-shot without RAG, ReAct enhances syntax success rates by 25.7\%, 26.4\%, and 31.2\% with iterative feedback from Simple, iverilog, and Quartus, respectively. We also observe consistent improvement from ReAct, regardless of the compiler and use of RAG.

\noindent\textbf{Impact of RAG:} Notably, the application of RAG with human expert guidance boosts the fix rate considerably and substantially enhances the solution's reliability. The results with Quartus compiler in Table~\ref{tab:compile} show that RAG improves the fix rate by 31.2\% (58.7\% to 89.9\% ) for One-shot and 18.6\% (79.9\% to 98.5\%) when using ReAct. We observe consistent improvement with RAG, regardless of the quality of the compiler feedback message and LLM (GPT-4).

\begin{table}[ht]
\begin{tabular}{ |l|cc| }
 %\hline
 %\multicolumn{4}{|c|}{VerilogEval-syntax} \\
 \hline
 LLM  & Syntax Success Rate & pass@1 \\
 \hline
 % GPT-3.5 & VCS & 55\%  \\ 
 GPT-3.5 &  73\% & 11\%  \\ 
 GPT-3.5 + \NAME & 93\% & 16\% \\
 \hline
\end{tabular}
\centering
\caption{Improvements of Syntax Success rate and simulation Pass@1 on RTLLM benchmark using ReAct and RAG with Quartus compiler. }
\label{tab:rtllm}
\end{table}

\noindent\textbf{Simulation Correctness Improvement:}
Previous research~\cite{liu2023verilogeval,lu2023rtllm} evaluates the performance of LLM-generated Verilog code using the pass@k metric. However, this approach doesn't account for syntax errors in the code samples, which can skew accuracy. In our study, we evaluate functional correctness using the VerilogEval benchmark, specifically addressing fixes to syntax errors in the code samples. The results, displayed in Table~\ref{tab:simulation}, show the performance scores on the VerilogEval dataset and the improvements after rectifying syntax errors, with 32.3\% and 10.1\% improvement on the pass@1 metric for Machine and Human respectively. We further divided the VerilogEval benchmark into two subsets: \textit{easy}, comprising 71 problems, and \textit{hard}, consisting of 85 problems. These subsets have been delineated based on a pass rate threshold of 0.1 on Human.
For simple problems in Human and low-level descriptions in the Machine, the correction of syntax errors significantly enhances the pass rate, reaching around 80\% for pass@1. When contrasting the pass rate improvements between easy and hard problems in the Human descriptions, we observe a greater improvement for easy problems at 14.5\% compared to hard problems at 6.7\% for pass@1. This discrepancy suggests that LLMs still face challenges when advanced reasoning and problem-solving skills are required. Discussions on future work to address simulation errors are presented in Section~\ref{sec:analysis}.

% {\color{blue} More results compare between easy and hard. Waiting for Machine results.}

\noindent\textbf{Generalizability:} Our method using ReAct and RAG can be generalized to other benchmark. 
%The VerilogEval-syntax dataset and our instructional prompts, derived from the VerilogEval benchmark, involve manually designed prompts. 
To account for potential overfitting during the design of the retrieval database, we also tested our method on the RTLLM~\cite{lu2023rtllm} benchmark without deriving new human guidance for the retrieval database. %Despite using different compilers, 
%Results in Table~\ref{tab:rtllm} confirm the effectiveness of our method, demonstrating its capability to generalize to unseen designs as well. 
As shown in Table~\ref{tab:rtllm}, our framework improves the syntax success rate from 73\% to 93\%, demonstrating its capability to generalize\footnote{The syntax success rate we collected is different from the original paper because we used the Verilog code sample provided in the their repo for each problem.}.

\subsection{Ablation Studies}
\label{sec:exp:ablation}
Our research includes two ablation studies designed to evaluate the impact of feedback quality and the selection of LLMs on the effectiveness of syntax error correction. %We assessed the performance under the One-Shot and ReAct.

\subsubsection{Impact of Feedback Quality}
\label{sec:ablation-feedback}
We study the impact of feedback quality with using different feedback messages detailed below.

\noindent\textbf{Simple:}
We only give an instruct prompt "\textit{Correct the syntax error in the code.}" without any explicit instruction on what the error is about and how to fix it.

\noindent\textbf{Icarus Verilog (iverilog)~\cite{williams2002icarus}:}
%In our study, we conducted a comparative analysis of two compilers: the open-source Icarus Verilog (iverilog)\cite{williams2002icarus} and the commercial (free edition) Quartus\footnote{https://www.intel.com/content/www/us/en/products/details/fpga/development-tools/quartus-prime/resource.html}. Figure~\ref{fig:compiler-feedback} showcases examples of compiler logs from each, highlighting their distinct characteristics.
%Iverilog, being open-source, tends to be less sophisticated. It occasionally encounters edge cases where it fails to provide informative compiler logs, sometimes merely outputting a message like ``I give up''. In many instances, its logs lack clarity, making them difficult to decipher except by experienced professionals, as further discussed in Section~\ref{sec:analysis}.
%On the other hand, Quartus, a commercial compiler, offers well-defined and clear compiler logs. It not only identifies errors effectively but often provides suggestions for rectification and additional validation tips, making it more user-friendly and informative for a broader range of users.
%In our comparative analysis of compilers, 
Open-source Verilog simulator. The compiler occasionally encounters edge cases where it fails to provide informative logs, outputting messages such as "\textit{I give up}." Logs lack clarity, making them challenging to decipher. 

\noindent\textbf{Quartus\footnote{https://www.intel.com/content/www/us/en/products/details/fpga/development-tools/quartus-prime/resource.html}:} 
Commercial compiler for FPGAs. In contrast with the open-source counterpart, it delivers well-defined and clear logs, effectively identifying errors and often offering suggestions and validation tips, making it more user-friendly and informative.

% ModelSim\footnote{https://eda.sw.siemens.com/en-US/ic/modelsim/}

%The first study, detailed in Table~\ref{tab:compile}, investigates the effect of using various compiler logs as feedback. We compared the outcomes using different compilers (iverilog, and Quartus).
% With One Shot prompting, the success rate of syntax error resolution improved significantly when using RAG, especially with Quartus (from 0.587 to 0.899).
% The ReAct prompting method showed a notable increase in error resolution effectiveness with Quartus when using RAG, achieving a remarkable success rate of 0.985.
%The results indicate that the clarity and informative of compiler logs  have substantial effects on the efficiency of the debugging process.

We deem Simple, iverilog, and Quartus to have increasing level of feedback message quality and illustrate the difference of the two compilers with an example in Figure~\ref{fig:compiler-feedback}. Results depicted in Table~\ref{tab:compile} clearly demonstrate that compiler logs give better feedback than Simple feedback and that the quality of compiler output impacts the performance of LLM debugging. As the quality of the compiler message improves (iverilog v.s. Quartus), the success rate of fixing syntax errors also increases. Intriguingly, the disparity between iverilog and Quartus results is more pronounced when using React with RAG. This discrepancy potentially suggests that high-quality compiler messages enhance the LLM's ability to more effectively utilize the retrieved human expert guidance.

\begin{figure}[h!]

\begin{center}

\begin{tcolorbox}[width=1.0\linewidth, halign=left, colframe=black, colback=white, boxsep=0.01mm, arc=1.5mm, left=2mm, right=2mm, boxrule=0.5pt]\footnotesize

\textbf{Task ID:} vector100r\\

\vspace{0.1in}
\textbf{Erroneous Implementation}\\  
\begin{minted}[breaklines]{verilog}
1  module top_module (
2  input [99:0] in,
3  output reg [99:0] out
4  );
5  always @(posedge clk) begin
6    for (int i = 0; i < 100; i = i +1) begin
7        out[i] <= in[99 - i];
8    end
9  end
10 endmodule
\end{minted}
\end{tcolorbox}

\begin{tcolorbox}[width=1.0\linewidth, halign=left, colframe=black, colback=white, boxsep=0.01mm, arc=1.5mm, left=2mm, right=2mm, boxrule=0.5pt]\footnotesize

\textbf{iverilog:}\\ 
vector100r.sv:5: error: Unable to bind wire/reg/memory `clk' in `top\_module'\\
vector100r.sv:5: error: Failed to evaluate event expression 'posedge clk'.\\
2 error(s) during elaboration.\\

\end{tcolorbox}

\begin{tcolorbox}[width=1.0\linewidth, halign=left, colframe=black, colback=white, boxsep=0.01mm, arc=1.5mm, left=2mm, right=2mm, boxrule=0.5pt]\footnotesize

\textbf{Quartus:}\\ 
Error (10161): Verilog HDL error at vector100r.sv(5): object "clk" is not declared. Verify the object name is correct. If the name is correct, declare the object. File: /tmp/tmp4u6ib9ig/vector100r.sv Line: 5
Error: Quartus Prime Analysis \& Synthesis was unsuccessful. 1 error, 1 warning
\end{tcolorbox}
\end{center}
\caption{Example of compiler log from iverilog and Quartus. Quartus feedback messages are more informative.}
\label{fig:compiler-feedback}

% \centering
% \includegraphics[width=\linewidth]{Figs/compiler-feedback.drawio.pdf}
\end{figure}

\subsubsection{Impact of Different LLMs}
\label{sec:ablation-llm}
In Table~\ref{tab:compile}, we present the results obtained when utilizing GPT-4 as the underlying LLM with Quartus as the compiler. A notable improvement in syntax error resolution is observed when using GPT-4 compared to GPT-3.5, particularly with One-shot prompting with RAG, where the success rate increased from 89.9\% to 98\%. When comparing the results of GPT-4 between One-shot and ReAct, we observe minor improvements of approximately 1\%, suggesting that GPT-4 is already a robust agent proficient in fixing syntax errors without the need for reasoning, action planning, and iterative refinement. 

Nonetheless, it is important to highlight that our approach of empowering LLMs with ReAct and RAG can significantly narrow the gap between weaker LLMs and stronger ones, especially beneficial for weaker open-source models~\cite{thakur2023verigen, liu2023verilogeval} that may not be as performant as GPT-4 on programming tasks.
%Our implementation leverages the OpenAI function call feature for formatting LLM responses, which currently limits its applicability to other open-source models. Expanding this evaluation to include more models is a prospective avenue for future research.

% \begin{table}[h]
% \centering
% \caption{Comparing the results using different LLM for fixing syntax error.}
% \label{tab:ablation-llm}
% \begin{tabular}{ |ll|cc| }
%  \hline
%  Prompt & Feedback & GPT-3.5 & GPT-4 \\
%  \hline
%  One Shot & w/o RAG & 0.57 & \textbf{0.91} \\ 
%  ~ & RAG & 0.899 & \textbf{0.97}  \\ 
%  % \hline
%  % ReAct & Compiler logs & 0.731 & \\ 
%  % ~ & RAG & \textbf{0.985} &  \\ 
%  \hline
% \end{tabular}
% \end{table}

\section{Analysis and Discussion}
\label{sec:analysis}
In this section, we delve into a series of analyses and discussions, extracting valuable insights from our discoveries. Specifically, we provide analysis in fail cases and the effect of iterative code refinement. Additionally, we discuss the challenges associated with applying our method to debugging simulation logic errors.

\noindent\textbf{Failure due to LLM's Incapability:}
Most of the cases where, even with the aid of ReAct and RAG, failed to correct syntax errors is due to the fundamental incapability of the LLM.
Figure~\ref{fig:analysis-failcase} illustrates one of the failure case where it particularly requires arithmetic index calculations to solve the index out-of-range error. Some other notable failures occurred in cases where LLMs were confident in incorrect syntax, possibly due to it being accepted in C/C++.

\begin{figure}[h]
\begin{tcolorbox}[width=1.0\linewidth, halign=left, colframe=black, colback=white, boxsep=0.01mm, arc=1.5mm, left=2mm, right=2mm, boxrule=0.5pt]\footnotesize

\textbf{Erroneous Implementation (Partial):}\\  
\begin{minted}[breaklines]{verilog}
for (i = 0; i < 16; i = i + 1) begin : ROW
    for (j = 0; j < 16; j = j + 1) begin : COLUMN
        neighbors[0] = q[(i-1)*16 + (j-1)];
        row_above = q[((i-1) & 15)*16 + j];
        ...
\end{minted}
%               neighbors[1] = q[(i-1)*16 + j];
%        row_below = q[((i+1) & 15)*16 + j];
\vspace{0.1in}
\textbf{Compile Error:}\\ 
Error (10232): Verilog HDL error at conwaylife.sv(23): index -17 cannot fall outside the declared range [255:0] for vector "q" 
%\\
%Error (12153): Can't elaborate top-level user hierarchy\\
%Error: Quartus Prime Analysis \& Synthesis was unsuccessful. 2 errors, 5 warnings\\

\end{tcolorbox}
\caption{An example which the agent failed to fix a syntax error. LLM failed to calculate array indices in the for loop and does not recognize the out-of-bound error.}
\label{fig:analysis-failcase}
\end{figure}

% \begin{figure}[h!]
% \centering
%   \centering
%   \includegraphics[width=\linewidth]{Figs/analysis-failcase.drawio.pdf}
%   \caption{Examples where RAG and ReAct failed to fix syntax error.}
%   \label{fig:analysis-failcase}
% \end{figure}

% \subsection{Easy problems benefits from fixing Syntax}

% Figure~\ref{fig:analysis-passrate} presents the improvement in pass@1 rates for each problem. The data indicates that rectifying syntax errors significantly boosts pass@k rates for simpler questions. However, for more complex problems, the approach seems to shift the errors from syntax to simulation.

% \begin{figure}[h!]
% \centering
%   \centering
%   \includegraphics[width=\linewidth]{example-image-a}
%   \caption{A bar chart on how many iteration needed to fix the compilation error.}
%   \label{fig:analysis-passrate}
% \end{figure}

\begin{figure}[h!]
\centering
  \centering
  \includegraphics[width=0.8\linewidth]{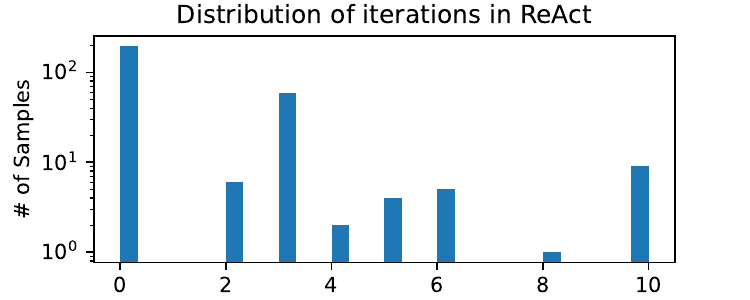}
  \caption{Distribution of iterations required by ReAct to fix syntax errors.}
  \label{fig:analysis-distribution}
\end{figure}

\noindent\textbf{Iterative Code Refinement:}
In Figure~\ref{fig:analysis-distribution}, we analyze the number of iterations ReAct requires to fix syntax errors. About 90\% of problems are resolved in a single revision. For the remaining cases, additional code revisions are necessary, as new errors may surface after addressing the initial ones.

\noindent\textbf{Challenges in Debugging Simulation Errors:}
While our framework could be readily adapted to employ LLMs for debugging simulation errors, our preliminary studies revealed limited improvements beyond syntax error fixes. Despite our efforts to provide simulation error logs as feedback to LLM agents, including summaries on output error count and text-formatted waveform-like comparisons of error versus solution output, we observed that LLMs had constrained capabilities to comprehend simulation feedback messages. They only exhibited proficiency in fixing logic implementation errors for simple problems but struggled with more complex questions, especially those involving high-level design functionality descriptions and advanced reasoning. Addressing the challenges associated with LLMs fixing erroneous implementations in such problems, particularly those requiring advanced reasoning and problem-solving skills, remains an exciting area for future research. This also highlights the need to improve LLM's capabilities in reasoning and problem-solving related to hardware design.

\section{Conclusion}
\label{conclusion}
Our framework \NAME~demonstrates the significant impact of employing Retrieval Augmented Generation (RAG) and advanced prompting methods like ReAct in debugging Verilog code with Large Language Models. Key findings indicate that these approaches notably enhance syntax error resolution, achieving success rates as high as 98.5\%. This research not only offers a novel autonomous language agent for Verilog code debugging but also introduces comprehensive dataset for further exploration.

% \begin{acks}
% \end{acks}
\bibliographystyle{ACM-Reference-Format}
\bibliography{sample-base}
% \appendix

\end{document}